\documentclass{sig-alternate-per}
\usepackage{algorithm,algpseudocode}

\newtheorem{definition}{Definition}

\usepackage{subfigure}
\usepackage{bm}
\usepackage{url}
\usepackage{times}
\usepackage{mathptmx}

\author{Linan Huang and Quanyan Zhu\\
{\{lh2328, qz494\}@nyu.edu}
\vspace{12pt}\\
Department of Electrical and Computer Engineering, New York University\\
 2 MetroTech Center, Brooklyn, NY, 11201, USA
\thanks
{This work was partially supported by NSF awards CNS-1544782, SES-1541164, ECCS-1550000, CNS-1720230, DOE grant DE-NE0008571, and a DHS grant through Critical Infrastructure Resilience Institute (CIRI).} 
 }
 
\title{Adaptive Strategic Cyber Defense for Advanced Persistent Threats in Critical Infrastructure Networks}

\begin{document}
\maketitle
\begin{abstract}
 Advanced Persistent Threats (APTs) have created new security challenges for critical infrastructures due to their stealthy, dynamic, and adaptive natures. In this work, we aim to lay a game-theoretic foundation by establishing a multi-stage Bayesian game framework to capture incomplete information of deceptive APTs and their multi-stage multi-phase movement. The analysis of the perfect Bayesian Nash equilibrium (PBNE) enables a prediction of attacker's behaviors and a design of defensive strategies that can deter the adversaries and mitigate the security risks.  A conjugate-prior method allows online computation of the belief and reduces Bayesian update into an iterative parameter update. The forwardly updated parameters are assimilated into the backward dynamic programming computation to characterize a computationally tractable and time-consistent equilibrium solution based on the expanded state space. The Tennessee Eastman (TE) process control problem is used as a case study to demonstrate the dynamic game under the information asymmetry and show that APTs tend to be stealthy and deceptive during their transitions in the cyber layer and behave aggressively when reaching the targeted physical plant. 
The online update of the belief allows the defender to learn the behavior of the attacker and choose strategic defensive actions that can thwart adversarial behaviors and mitigate APTs. 
Numerical results illustrate the defender's tradeoff between the immediate reward and the future expectation as well as the attacker's goal to reach an advantageous system state while making the defender form a positive belief.

\end{abstract}

\section{Introduction}
\begin{figure}[]
  \centering
  \includegraphics[width=1\columnwidth]{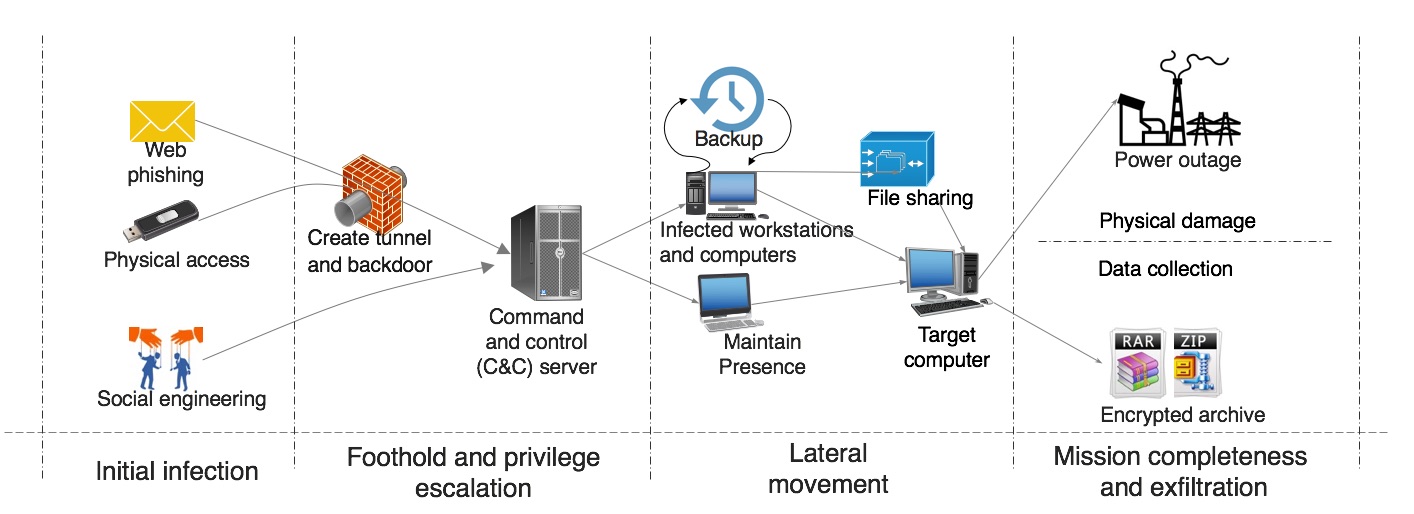}
  \caption{\label{attack graph} 
  APTs start the infection by exploiting network vulnerabilities or the human weakness. They aim to cause physical damages or collect confidential data. 
  }
\end{figure}

With the integration of communication networks and information technologies with the critical infrastructures including power grids, transportation systems, and water distribution systems, the direct use of the off-the-shelf technologies has made our infrastructure vulnerable to cyber attacks. One emerging threat is the Advanced Persistent Threats (APTs) which are a class of multi-phase and multi-stage hacking processes \cite{zhu2018multi}, initiating their infections in cyberinfrastructures yet targeting at specific physical infrastructures such as nuclear power stations and automated factories.  
Unlike the ``spray-and-pray" attacks, APTs as the targeted attacks, perform reconnaissance and tailor their hacking techniques to the targeted system. 
As shown in Fig. \ref{attack graph}, the APTs' life cycle includes a sequence of phases and stages such as the initial entry, privilege escalations, and lateral movements. APTs use each stage as a stepping stone for the next one. Since APTs have a specific target at the final stage, they receive no benefits going back to previous stages. Thus, the multi-stage attack graph bears a tree structure without jumps or loops.
Unlike the ``smash-and-grab" attacks, APTs behave seemingly as legitimate users, wait until the final stage to launch the  ``critical hit'' on their specific targets, and inflict an enormous loss. 

The classical intrusion prevention techniques such as the cryptography and the physical isolation can be ineffective for APTs. An APT-type adversary can steal the full cryptographic keys by exploiting zero-day vulnerabilities and techniques such as social engineering. Stuxnet can bridge the air gap between local-area networks with the insertion of infected USB drives.
Similarly, the intrusion detection approach \cite{coppolino2010intrusion} has proven to be insufficient when APTs acquire knowledge of the system response as well as the detection rule with the help of insiders and the reconnaissance. 
Moreover, APTs operated by human experts can analyze, learn, and update the knowledge of the system, thus evading detection by stealthy and strategic movements, e.g., scan the port sufficient slow to avoid the alarm and even choose the No Operation (NOP) at some stages.
  Hence, it is essential to design up-to-date security mechanisms that can mitigate the risks despite the successful infiltration and the strategic response of APTs.

One way to understand the multi-stage and stealthy nature of the APTs is through dynamic games with incomplete information. The dynamic game frameworks capture the multi-stage movement of the defender and the attacker in networks \cite{Huang2017,manshaei2013game}. The deceptive and stealthy behaviors of the APTs lead to the information asymmetry where 
an attacker has his own private information encapsulated by a random variable called  \textit{types} \cite{harsanyi1967games}. The \textit{type} characterizes the essence and the objective of the user, i.e., whether the user is legitimate or adversarial, which assets serve as his targets, and how much damages he can inflict on the system. 
The user's type determines his behaviors if he is rational and aims at maximizing his utility, which makes it possible for the defender to form and update a belief of the type based on the history of user's behaviors. 
Since the attacker has to follow the network protocol and move stealthily across the networks by hiding his footprints and evading the detection, it is natural to view the defender as the principal who can design security policies and the attacker as the agent who follows the policies to attain his goal. 
The strategic behaviors of the defender and the attacker will lead to a perfect Bayesian Nash equilibrium (PBNE) where no one can profit from unilateral deviations  at any stage. 
Achieving a long-term statistic optimal is challenging since the belief updates forwardly yet the PBNE strategy pair is computed backwardly. 
With the beta-binomial conjugate prior assumption, we manage to unify the coupled forward and backward processes and form the dynamic programming with an expanded state. 
 Tennessee Eastman process is used as a case study to illustrate the theoretical underpinning of our framework for the design of strategic defense to deter the attacks and mitigate the impact of the threats.
\\
\textbf{Related Work}:
\texttt{FlipIt} game \cite{van2013flipit} has analyzed the scenario of the key leakage under APTs so that a system operator and APTs will takeover the system alternately. Defenders cannot know the time of the stealthy takeover as well as the current system status unless taking defensive actions. \texttt{FlipIt} game provides high-level guidelines on how to allocate the limited resources to deter the APTs. Our multi-stage Bayesian game framework, however, supports a specification of both adversarial and defensive actions with utilities and enables the equilibrium analysis of the game as the prediction of the attack moves. 
Signaling game, a two-stage game with the one-sided \textit{type}, has been applied to study the information asymmetry in cyber deception \cite{zhang2017strategic}. However, both players receive a one-shot utility which does not well capture the multi-stage transition of the APTs. In our framework, each player at each stage receives feedbacks involving his/her immediate reward and the other player's apparent activities, which enables the defender to learn the attacker's type during the multi-stage interactions. 
\\
\textbf{Organization of the Paper}:
The rest of the paper is organized as follows. Section \ref{modelSection} introduces the forward belief update and the backward dynamic programming under the PBNE solution concept.
In Section \ref{ComputationSection}, we adopt the binomial-beta {conjugate prior} to turn the nonparametric update of the distribution into a parametric one. A case study of APTs targeted at the TE process is presented in Section \ref{casestudySection}, and Section \ref{ConclusionSec} concludes the paper. 
\section{System Model }\label{modelSection}
Consider a two-person game with $P_1$ as the system defender  (pronoun 	``she'') and $P_2$ as the user (pronoun ``he'').
The user has a type $\theta$ which is the realization of a continuous random variable $\tilde{\theta}$  with the support $\Theta:=[0,1]\subset \mathbb{R}$. 
 The value of the type indicates the strength of the user in terms of damages that he can inflict on the system. A user with a larger type value indicates a higher threat level to the system. 

At each stage $t\in \{0,1,\cdots,T\}$, each player $P_i$ chooses an action $a_i^t\in \mathcal{A}_i^t$. The user's actions represent the apparent behaviors and observable activities from log files such as a privilege escalation request and sensor access. A defender cannot identify the user's type from observing his actions.  
The defender's action $a_1^t$ represent precautions and proactive behaviors such as restricting the escalation request or monitoring the sensor access. 
 Thus, the action pair $(a_1^t,a_2^t)$ is known to both players after stage $t$ and forms a \textit{history} ${h}^t:=\{a_1^0,\cdots,a_1^{t-1},a_2^0,\cdots,a_2^{t-1}\} $. 
The state $x^t\in \mathcal{X}^t$ shows the system status such as the location of the APTs at each stage $t\in \{0,1,\cdots,T\}$.  Since the initial state $x^0$ and history $h^t$ uniquely determine the state,  $x^t$ contains information of history up to $t$ and has the transition kernel described by $x^{t+1}=f^t(x^t , a_1^t,a_2^t )$ with a  deterministic kernel  function $f^t$.
Define  $\bigtriangleup \mathcal{A}_i^t$ as the probability distribution over $P_i$'s action space.
The \textit{behavioral} mixed strategies $\sigma^t_1: \mathcal{X}^t  \mapsto \bigtriangleup \mathcal{A}_1^t $ and $\sigma^t_2: \mathcal{X}^t \times \Theta \mapsto \bigtriangleup \mathcal{A}_2^t $ mean that both players make their decisions based on the information available to them. With a slight abuse of notation, let $\sigma^t_1(a_1^t|x^t),\sigma^t_2(a_2^t|x^t, \theta)$ be the probability of taking action $a_i^t$ at stage $t$ under state $x^t$ and type $\theta$.  
The set of all behavioral mixed strategies $\sigma_i^t$ forms the strategy space $\Sigma_i^t$.
\\
\textbf{Believe Update:}
To strategically gauge the user's type, the defender specifies a belief $B^t: \mathcal{X}^t  \mapsto \bigtriangleup \Theta$ as a distribution over the type space according to the state $x^t$ at stage $t$. Likewise, $B^t(\theta|{x}^t)$ is the conditional probability density function (PDF) of the type $\theta$ and $\int_0^1 B^t(\theta|{x}^t)d\theta=1,  \forall t, {x}^t$.  
The prior distribution of the user's type is known to be $B^0$ and the belief of the type updates according to the Bayesian rule with the arrival of the action observation $a_2^t$ drawn from the mixed strategy $\sigma_2^t$. 
\begin{equation}
\label{eq: forward}
B^{t+1}(\theta|f^t({x}^t,{a}_1^t,{a}^t_{2}) )=\frac{B^t(\theta|{x}^t)\sigma^t_{2}({a}_{2}^t|{x}^t,\theta)}{\int_0^1 B^t(\hat{\theta}|{x}^t)\sigma^t_{2}({a}_{2}^t|{x}^t,\hat{\theta})d\hat{\theta}}.
\end{equation}
\\
\textbf{Utility Function}:
The user's type influences $P_i$'s immediate payoff received at each stage $t$, i.e., $J_i^t: \mathcal{X}^t \times \mathcal{A}_1^t\times \mathcal{A}_2^t \times \Theta \mapsto \mathbb{R}$. For example, a legitimate user's access to the sensor benefits the system while a pernicious user's access can incur a considerable loss. 
Define $\sigma_i^{t':T}:=\{\sigma^t_i \in \Sigma_i^t\}_{t=t',\cdots,T}\in \Sigma_i^{t':T}$ as a sequence of policies from $t'$ to $T$. The defender has the objective to maximize the cumulative expected utility: 
\begin{equation*}
\begin{split}
&U^{t':T}_1(\sigma_1^{t':T},\sigma_{2}^{t':T}, {x}^{t'}):= \sum_{t=t'}^{T} E_{\theta \sim B^t , a_{1}^t\sim \sigma_1^t, a_2^t\sim \sigma_2^t } J_1^t(x^t,a_1^t,a_{2}^t,\theta)\\
&= \sum_{t=t'}^{T}  \int_0^1 B^t(\theta |{x}^t)  \sum_{a_{1}^t \in \mathcal{A}_1^t} \sigma_{1}^t(a_{1}^t |{x}^t)
\sum_{a_{2}^t\in \mathcal{A}_2^t }\sigma_{2}^t(a_{2}^t |{x}^t, \theta)      J_1^t  d\theta.   
\end{split}
\end{equation*}
and the user's objective function is 
\begin{equation*}
\begin{split}
&U^{t':T}_2(\sigma_1^{t':T},\sigma_{2}^{t':T}, {x}^{t'},\theta)= \sum_{t=t'}^{T}    \sum_{a_{1}^t \in \mathcal{A}_1^t} \sigma_{1}^t(a_{1}^t |{x}^t)
\sum_{a_{2}^t\in \mathcal{A}_2^t }\sigma_{2}^t(a_{2}^t |{x}^t, \theta)      J_2^t  .   
\end{split}
\end{equation*}
\\
\textbf{Perfect Bayesian Nash Equilibrium}:
We model the scenario of APTs under the insider threat as a dynamic principal-agent problem  where defender $P_1$ as the principal chooses her policy $\sigma_1^t$ first at each stage $t$. Attacker $P_2$ as the agent perceives $\sigma_1^t$ via insiders, and then chooses his policy $\sigma_2^t$ to best-respond to $\sigma_1^t$, i.e.,  maximizes his cumulative expected utility $U_2^{t:T}$. 
Since APTs have to follow rules to avoid detection, a sophisticated defender aware of the potential policy leakage under insider threats can acquire the best response of APTs through the attack tree or honeypots. 
The described security scenario leads to the following definition of perfect Bayesian Nash equilibrium (PBNE) where the defender chooses the most rewarding policy to confront the attacker's best-response policies. 

\begin{definition}
\label{defin: best response}
In the two-person multi-stage game with a sequence of beliefs $B^t, t\in \{t',\cdots,T\}$ satisfying the Bayesian update in \eqref{eq: forward} and the cumulative utility function $U_i^{t':T}$, the set $R_2^{\theta,x^{t'}}(\sigma_1^{t':T}):=\{\gamma \in \Sigma_2^{t':T}: U_2^{t':T}(\sigma_1^{t':T},\gamma,x^{t'}, \theta)\geq U_2^{t':T}(\sigma_1^{t':T},\sigma_2^{t':T},x^{t'}, \theta), \forall \sigma_2^{t':T}\in \Sigma_2^{t':T}, \forall x^{t'}\in \mathcal{X}^{t'}, \theta\in \Theta\}$
is $P_2$'s \textbf{best-response set} to $P_1$'s policy $\sigma_1^{t':T} \in \Sigma_1^{t':T}$ under state $x^{t'}$ and type $\theta$.
\qed
\end{definition}

\begin{definition}
\label{defin: P-PBNE}
In the two-person multi-stage Bayesian game with $P_1$ as the principal, the cumulative utility function $U_i^{t':T}$, the initial state $x^{t'}\in \mathcal{X}^{t'}$, the type $\theta \in \Theta$, and a sequence of beliefs $B_i^t, t\in \{t',\cdots,T\}$ in \eqref{eq: forward}, a sequence of strategies $\sigma_1^{*,t':T}\in \Sigma_1^{t':T}$ is called a perfect Bayesian Nash equilibrium (\textbf{PBNE}) for the principal, if 
\begin{equation*}
\begin{split}
U_1^{*,t':T}(x^{t'}):=&\inf_{\sigma_2^{t':T}\in R^{\theta,x^{t'}}_2(\sigma_1^{*,t':T})} U_1^{t':T}(\sigma_1^{*,t':T},\sigma_{2}^{t':T},x^{t'})\\
=&\sup_{\sigma_1^{t':T}\in \Sigma_1^{t':T}}\inf_{\sigma_2^{t':T} \in R^{\theta,x^{t'}}_2(\sigma_1^{t':T})} U_1^{t':T}(\sigma_1^{t':T},\sigma_{2}^{t':T},x^{t'}).
\end{split}
\end{equation*}
A strategy $\sigma_2^{*,t':T}\in arg\max_{\sigma_2^{t':T}\in \Sigma_2^{t':T}} U_2^{t':T}(\sigma_1^{*,t':T},\sigma_{2}^{t':T},x^{t'}, \theta)$ is a PBNE for the agent $P_2$.
\qed
\end{definition}
\textbf{Dynamic Programming}: \label{DPsection}
Given the type belief at every stage, we use dynamic programming to find the PBNE policies in a backward fashion because of the tree structure and the finite horizon. 
Define the value function $V_1^{t}({x}^{t}):=U_1^{t: T}(\sigma_1^{*,t:T},\sigma_{2}^{*, t:T}, x^t )$ and $V_2^{t}({x}^{t},\theta):=U_2^{t: T}(\sigma_1^{*,t:T},\sigma_{2}^{*, t:T}, x^t ,\theta)$ as the optimal utility-to-go for the defender and the user, respectively. 
We have the following simultaneous equations, i.e., 
\begin{equation*}
\begin{split}
&V_1^{t}({x}^{t})= \sup_{\sigma_1^{t}}  
E_{\theta \sim B^{t}, a_1^{t}\sim \sigma_1^{t},a_2^{t}\sim \sigma_{2}^{*,t}} [ V_1^{t+1}(f^t(x^t,a_1^t,a_2^t)) 
+  J_1^{t}(x^{t}, a_1^{t},a_{2}^{t})],\\
&V_2^{t}({x}^{t},\theta)=  \sup_{\sigma_2^{t}} 
E_{a_2^{t}\sim \sigma_2^{t},a_1^{t}\sim \sigma_{1}^{*,t}} [ V_2^{t+1}(x^{t+1},\theta) 
+  J_2^{t}(x^{t}, a_1^{t},a_{2}^{t},\theta)], 
\end{split}
\end{equation*}
where $\sigma_i^{*,t}, i\in \{1,2\}$ is the PBNE policy pair at stage $t$. 
The above system equations have to be solved backwardly from stage $t$ to stage $0$ with the boundary conditions $V_1^{T+1}({x}^{T+1}),V_2^{T+1}({x}^{T+1},\theta)$ at stage $t=T+1$. However, the belief $B^t$ in \eqref{eq: forward} updates forwardly with the boundary condition $B^0$ at initial stage $t=0$. These two equations are coupled, and we need to find the consistent pair of PBNE strategies and beliefs.
 
\section{Computation}\label{ComputationSection}
In the Bayesian update, the prior probability distribution $B^{t}$ is called a conjugate prior for the likelihood function $\sigma^t_{2}$ if the posterior distribution $B^{t+1}$ is in the same family as the prior distribution $B^{t}$.  
Similar to our previous work \cite{huang2018gamesec}, the defender divides the action space of the user into $K+1$ time-invariant sets $ \mathcal{C}_j$, i.e., $\mathcal{A}_2^t=\{\cup \mathcal{C}_j\}_{j=0,1, \cdots, K},\forall t$ which are mutual exclusive $\mathcal{C}_j \cap \mathcal{C}_l = \emptyset, \forall  j\neq l $. 
Each set represents a category and each $a_{2}^t$ uniquely corresponds to one category. Then, we can transform $\sigma^t_{2}(a_{2}^t|{x}^t,\theta)$, the distribution of $a_2^t$, into a distribution of the corresponding category $\hat{\sigma}^t_{2}(k^t|{x}^t,\theta)$ with the index $k^t\in \{0,1,\cdots, K\}$. 
If we assume $\hat{\sigma}^t_{2}$ to be a binomial distribution with the parameter $q=\theta$ and $N=K$. The probability mass function (PMF) of category $k$ is
$\Pr(k)=\binom{N}{k}q^k (1-q)^{N-k}.$
The prior belief $B^0$ is assumed to be a beta-distribution with hyperparameters $\alpha^0$ and $\beta^0$. 
Since binomial and beta-distributions are conjugate, the posterior belief conserves to be a beta-distribution with updated hyperparameters $(\alpha^{t+1},\beta^{t+1})=(\alpha^t+k^t,\beta^t+K-k^t)$, where $k^t$ is the category that the user's action at stage $t$ falls into. 
Finally, we  transform the belief conditioned on the categories back to the belief conditioned on the corresponding actions using the hard de-aggregation in which actions $a_{2}^t, \bar{a}_{2}^t$ correspond to the same category $k^t$ share the same belief distribution $B^t$. 
\\
\textbf{Expanded State in Dynamic Programming}: 
Since the parameter update $(\alpha^{t+1},\beta^{t+1})=(\alpha^t+k^t,\beta^t+K-k^t)$ is sufficient to determine the belief update in \eqref{eq: forward}.
The original system state $x^t$ and the belief state $\alpha^t,\beta^t$ compose an expanded state ${y}^t=\{x^t,\alpha^t,\beta^t\}$. Since $\alpha^t+\beta^t=\alpha^0+\beta^0+tK$, we only need one of the two parameters to determine the beta-distribution and the notation $\theta \sim \beta^t$ means that the type is of the beta-distribution with the hyperparameters $(\alpha^0+\beta^0+tK-\beta^t, \beta^t)$. 
\begin{equation}
\label{eq: backward}
\begin{split}
&{V}_1^{t}(y^{t})= \sup_{{\sigma}_1^{t}\in \Sigma_1^t}  
E_{\theta \sim \beta^{t}, a_1^{t}\sim {\sigma}_1^{t},a_2^{t}\sim {\sigma}_{2}^{*,t}} [ {V}_1^{t+1}(y^{t+1}) 
+  J_1^{t}(x^{t}, a_1^{t},a_{2}^{t},\theta)], \\
&{V}_2^{t}(y^{t},\theta)=  \sup_{{\sigma}_2^{t}\in \Sigma_2^t} 
E_{a_2^{t}\sim {\sigma}_2^{t},a_1^{t}\sim {\sigma}_{1}^{*,t}} [ {V}_2^{t+1}(y^{t+1},\theta) 
+  J_2^{t}(x^{t}, a_1^{t},a_{2}^{t},\theta)]. \\
\end{split}
\end{equation}
Since  $\alpha^t=\alpha^0+\sum_{t'=1}^t k^{t'}, \beta^t=\beta^0+t K-\sum_{t'=1}^t k^{t'}, \forall t\in \{1,\cdots,T\}$, the number of feasible expanded states at stage $t$ is finite. Thus, we can directly compute \eqref{eq: backward} from stage $t=T$ to stage $t=0$ in a backward fashion and obtain the consistent pair of beliefs and PBNE policies. 

\section{Case Study}\label{casestudySection}
\begin{figure}[h]
\centering     
\includegraphics[width=1\columnwidth]{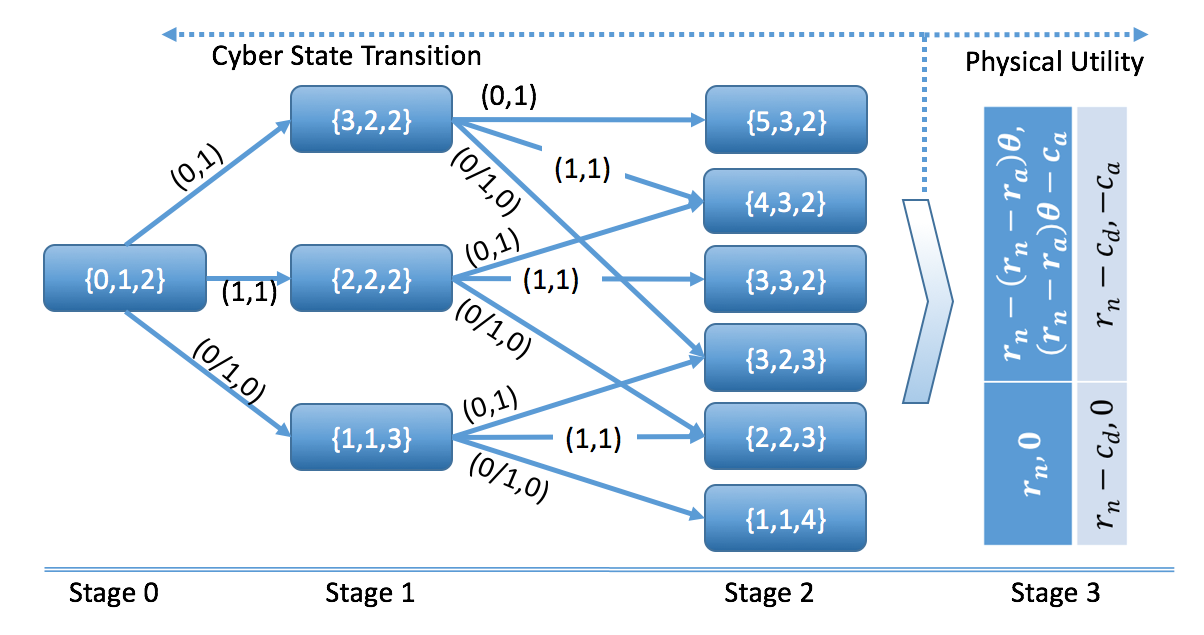}
\caption{\label{fig: TransitionMap} The multistage transition of the expanded state $y^t=\{x^t,\alpha^t,\beta^t\}$. 
The parenthesis around the arrow, e.g., $(0/1,1)$ denotes the value of the actions pair $(a_1^t,a_2^t)$.
}
\end{figure}

We consider a four-stage transition ($T=2$) of the expanded state $y^t=\{x^t,\alpha^t,\beta^t\}$ as shown in Fig. \ref{fig: TransitionMap}. At the initial stage $t=0$ where no behaviors are observed, the defender forms a biased initial belief that the user is more likely to be legitimate and of small threats, i.e., $B^0$ is a beta-distribution with hyperparameter $\alpha^0=1,\beta^0=2$. 
Starting from the initial system state $x^0=0$, the user at stage $t=0$ (and $t=1$) chooses to escalate his privilege $a_2^t=1$ (resp. propagate laterally) with an action cost $c_2^t\geq 0$ or no operation performed (NOP) $a_2^t=0$. The defender, at stage $t=0$ (and $t=1$), can choose proactive actions such as restricting the privilege escalation $a_1^t=1$  (resp. the lateral movement) with an action cost $c_1^t\geq 0$ or no operation performed (NOP) $a_1^t=0$. 
State $x^1=3,2,1$ represents a high, medium, and low privilege level for the user, respectively. The state transition function $f^0$ shows that if the user escalates his privilege $a_2^t=1$ and the defender does not restrict it $a_1^t=0$, the output privilege level is high; if the defender restricts it $a_1^t=1$, the output level is medium; otherwise the output level is low when the user takes NOP $a_2^t=0$.  
Let $K=1$, then the secure category $k=0$ includes $a_2^t=0$ and $k=1$ includes $a_2^t=1$, respectively. 

\begin{figure}[h]
\centering     
\includegraphics[width=1\columnwidth]{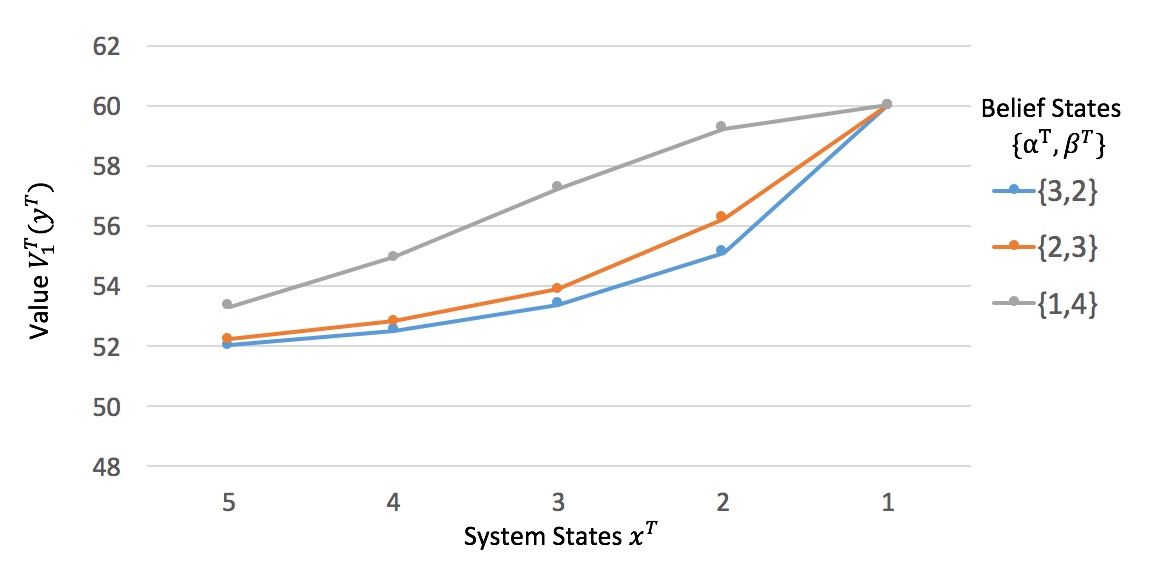}
\caption{ \label{fig: valueState} 
Defender's value function $V_1^T(y^T)$ under different expanded states $y^T=\{x^T,\alpha^T,\beta^T\}$.
}
\end{figure}

The traditional cyber security concerns the information protection yet APTs go beyond that. APTs can break the normal industrial operation by falsifying the set point of the controller, tampering the sensor reading and blocking the communication channel to cause delays in either the control message or the sensing data. 
Thus, after the transition in the cyberinfrastructure, the user will arrive at the physical plant (i.e., stage $t=2$) with the system state $x^t\in \mathcal{X}^t=:\{1,2,3,4,5\}$ representing different sensors under user's control. 
Both players take actions, obtain utilities relating to the  operation of the physical plant, and arrive at the terminal stage $t=T+1$ with the boundary conditions $V_1^{T+1}({x}^{T+1}):=0, V_2^{T+1}({x}^{T+1},\theta):=0, \forall \theta\in \Theta$, i.e., terminal states share the same stage utilities after the breach has happened. 
\\
\textbf{Physical Threats}:
We consider the benchmark Tennessee Eastman (TE) process as the targeted physical plant. 
The TE  process involves two irreversible reactions to produce two liquid (liq) products $G,H$ from four gaseous (g) reactants $A,C,D,E$.
\begin{align*}
A(g)+C(g)+D(g) \rightarrow G(liq),\\
A(g)+C(g)+E(g) \rightarrow H(liq).
\end{align*}
The process shuts down when the safety constraints are violated such as a high reactor pressure, a high/low separator/stripper liquid level.
The control objective is to maintain a desired production rate as well as quality, while stabilizing the whole system under Gaussian noise to avoid violating the safety constraints.
The inherent feedback controller for this nonlinear system performs well and results in the utility $r_n$ derived from three performance metrics, i.e., the product quality, the operation cost, and the shutdown time. 
Attackers can compromise different sensors and lead to different  states $x^T\in \mathcal{X}^T$. Then attackers can revise the reading to drive the system away from the reference point. Define a reward function $r_a: \mathcal{X}^T\mapsto \mathbb{R}$, then $r_a(x^T)$ will be the operation utility of the TE process under the state $x^T$, which can be obtained from the simulation results of the process model \cite{bathelt2015revision}. 
We rank the output value of the function $r_a$ from high to low and index the states correspondingly, e.g., $x^T=1$ indicates the compromise of a secondary sensor and $x^T=5$ indicates the compromise of all the sensors in the TE process. 
Action $a_2^T=0$ means NOP and $a_2^T=1$ means revising the readings of the sensors under his control. Unlike the stealthy transition in the previous cyber networks, the attacker at the final stage $T$ do not need to disguise as legitimate and can take detectable adversarial actions. 
Defenders can choose to defend $a_1^T=1$ with the cost $c_1^T$ or not defend $a_1^T=0$. 
\begin{table}[t]
\centering
\caption{The nonzero-sum stage utilities $(J_1^T,J_2^T)$. }
\label{table: laststage}
\begin{tabular}{|c|c|c|}
\hline
 $(J_1^T,J_2^T)$ & $a^T_2=0$        & $a^T_2=1$          \\ \hline
$a^T_1=0 $    & $(r_n, 0)$       & $(r_n-(r_n-r_a)\theta, (r_n-r_a)\theta-c^T_2)$       \\ \hline
$a^T_1=1$      & $(r_n-c^T_1, 0 )$ & $(r_{n}-c^T_1, -c^T_2 )$ \\ \hline
\end{tabular}
\end{table}
As shown in Table \ref{table: laststage}, if $a_i^T=0, i\in \{1,2\}$, the system operates normally with a reward of $r_n$ and the user does not receive rewards incurred by the attack. If the action pair is $(1,0)$, the defender has to pay an extra cost to monitor the sensor activities. If both players take actions $(1,1)$, then the system is well protected and receives a normal operation utility minus the monitoring cost while the attack pays the action cost $c_2^T$ without accomplishing the compromise. 
Finally, if the attacker launches an attack under no proper defenses, the system is compromised and receives a discounted payoff $r_n-(r_n-r_a(x^T))\theta$. The attacker, on the other hand, wins a reward proportional to $(r_n-r_a(x^T))\theta$. 
 Note that the reward loss is discounted by the threat level $\theta$, At the extreme case $\theta=0$ where the legitimate user will not sabotage, i.e., $r_n-(r_n-r_a(x^T))\theta=r_n$ and he receives no benefits from the attack. 
Let $p^t(y^t),q^t(y^t,\theta)$ be the probability of taking action $a_1^t=1,a_2^t=1$, respectively. Thus, the value functions under the PBNE mixed-strategies are given as follows.
\begin{equation}
\label{eq: final stage}
\begin{split}
 V^T_1(y^T)= \max_{p^T}  \ \  E_{\theta\sim \beta^T}& [q^T(y^T,\theta) (r_n-r_a(x^T)) \theta-c_1^T p^T(y^T)\\
&+ r_n-q^T(y^T, \theta)  (r_n-r_a(x^T) ) \theta ],\\
V^T_2(y^T,\theta)= \max_{q^T(y^T,\theta)}  \ \  & [ (1-p^T(y^T))(r_n-r_a(x^T) ) \theta-c_2^T]q^T(y^T,\theta).
\end{split}
\end{equation} 
The best-response policy 
$
R_2^{\theta,x^T}(p^T)=\mathbf{1}_{\{(1-p^T)(r_n-r_a(x^T) ) \theta-c_2^T >0\}}=\mathbf{1}_{\{\theta>\bar{\theta}(p^T,x^T)\}},
$
 where $\bar{\theta}(p^T,x^T):=c_2^T/[(1-p^T)(r_n-r_2(x^T) )]$. 
Plug the best response into the first equation, and we solve for the defender's policy $p^{*,T}(y^T)$ as well as the user's policy $q^{*,T}(y^T, \theta)=R_2^{\theta,x^T}(p^{*,T}(y^T))$. The user's policy $q^{*,T}$ has the threshold $\bar{\theta}(p^T,x^T)$, i.e., if his type value $\theta<\bar{\theta}$, he will choose NOP; otherwise he will choose to attack. 
The policy $q^{*,T}(y^T, \theta)$ is \textit{semi-separating} if the threshold $\bar{\theta}\in (0,1)$ and is called a \textit{pooling strategy} if the threshold $\bar{\theta}\not \in (0,1)$. The defender cannot learn any knowledge about user's type when he adopts pooling strategies which are independent of his type value.
\\
\textbf{Cyber Transitions}: 
According to \eqref{eq: backward}, the value functions $V_i^t$ from stage $t=0,1$ under the PBNE can be computed in the same fashion as in \eqref{eq: final stage} if the future expectation $V_i^{t+1}$ is assimilated into the direct stage reward $J_i^t$ to form an equivalent stage utility $\tilde{J}_1^t(x^t,a_1^t,a_2^t, \theta)=V_1^{t+1}(y^{t+1})+J_1^t(x^t,a_1^t,a_2^t, \theta)$ and $\tilde{J}_2^t(x^t,a_1^t,a_2^t, \theta)=V_2^{t+1}(y^{t+1}, \theta)+J_2^t(x^t,a_1^t,a_2^t, \theta)$ as shown in Table \ref{costtable}. 
Since the attacker aims to compromise sensors and inflict physical damages at stage $T$, we assume a petty utility for the cyber state transition, i.e., only the action cost $c_i^t, i\in \{1,2\}$ is taken into account. 
However, actions at the cyber stage $t=0,1$ will affect the future system state $x^T$ at the physical stage. Thus, the defender has the tradeoff of being secure and economical, i.e., paying the defense cost to guard against the potential compromises. 
On the other hand, the attacker's action will also affect the future belief state $\alpha^T,\beta^T$, which leads to his tradeoff of either being stealthy or reaching advantageous future expanded states $y^T$. 
\begin{table}[]
\centering
\caption{ Equivalent stage utility $\tilde{J}_1^t$ and $\tilde{J}_2^t$ under the expanded state $y^t=\{3,2,2\}$ at stage $t=1$. }
\label{costtable}
\begin{tabular}{|c|c|c|}
\hline
$\tilde{J}_1^t$ & $a^t_2=0$        & $a^t_2=1$          \\ \hline
$a^t_1=0 $    & $V^{t+1}_1(\{3,2,3\})$       & $V^{t+1}_1(\{5,3,2\})$       \\ \hline
$a^t_1=1$      & $V^{t+1}_1(\{3,2,3\})-c_1^t$ & $V^{t+1}_1(\{4,3,2\})-c_1^t$ \\ \hline
\end{tabular}
\begin{tabular}{|c|c|c|}
\hline
$\tilde{J}_2^t$ & $a^t_2=0$        & $a^t_2=1$          \\ \hline
$a^t_1=0 $    & $V^{t+1}_2(\{3,2,3\}, \theta)$       & $V^{t+1}_2(\{5,3,2\}, \theta)-c^t_2$       \\ \hline
$a^t_1=1$      & $V^{t+1}_2(\{3,2,3\}, \theta)$ & $V^{t+1}_2(\{4,3,2\}, \theta)-c_2^t$ \\ \hline
\end{tabular}
\end{table}
\\
\textbf{Comparisons and Insights}: 
As shown in Fig. \ref{fig: valueState}, a high value for the defender is the result of a healthy system state $x^T$ as well as a belief state where the user is more likely to have a low type value.
At the extreme state $x^T=1$ where the reward incurred by the attack is so low that users with any type values choose not to attack. Then the defender does not need to defend $p^{*, T}(y^T)=0, \forall y^T$ and obtains the maximum utility.  

\begin{figure}[h]
\centering     
\includegraphics[width=1\columnwidth]{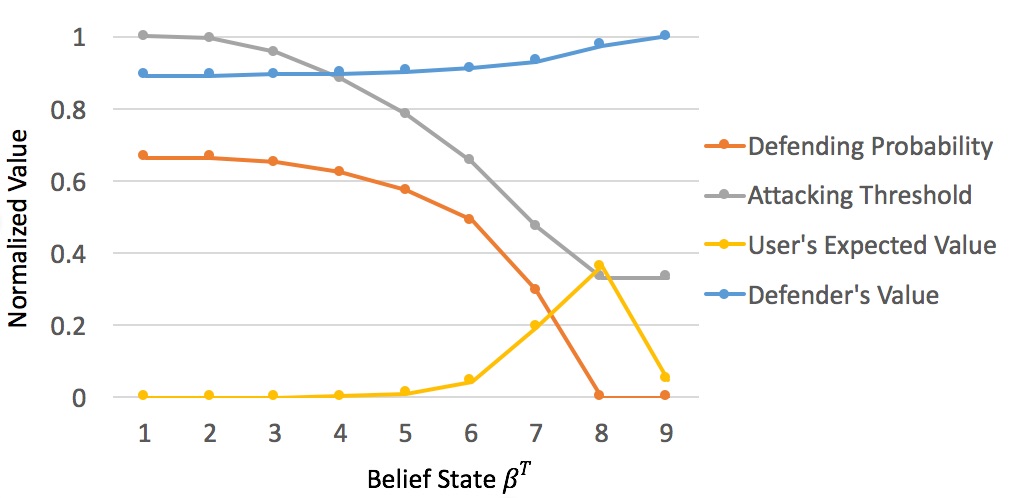}
\caption{ \label{fig: normalizedTrend} 
The effect of the defender's belief. 
}
\end{figure}

To investigate the effect of the defender's belief, we fix the system state $x^T=3$ and change the belief state $(\alpha^T, \beta^T)$ from $(9,1)$ to $(1,9)$, which means that the defender grows optimistically that the user is of a low threat level with a high probability. 
 Since players' value functions are of different scales in terms of the attacking threshold and the probability, we normalize the value functions with respect to their maximum values to illustrate their trends and make them comparable to the threshold and the probability as shown in Fig. \ref{fig: normalizedTrend}. 
 When $\beta^T$ is small, the defender chooses to protect the system with a high probability $p^{*,T}=0.67$, which completely deters attackers with any type values because the probability to attack $q^{*,T}=\mathbf{1}_{\{\theta>\bar{\theta}\}}$ equals $0$ when the attacking threshold $\bar{\theta}$ is $1$. 
 
As the defender trusts more about the user's legitimacy, the defending probability $p^{*,T}(y^T)$ decreases to $0$ when $\beta^T=9$.  
Since the defender is less likely to defend, the attacker bears a smaller threshold to launch the attack. 
However, the threshold will not decrease to $0$ because the users with type values less than $\bar{\theta}=0.33$ (defined as the limiting threshold) cannot receive sufficient rewards from the attack even when the defender chooses NOP. The value of the limiting threshold depends on the expanded state $y^T$, yet it should always be larger than $0$ because a user with type $\theta=0$  has no incentive to attack. 
The resulted defending policy $p^{*,T}$ captures a tradeoff between security and economy and guarantees a high value for defenders at most of the belief states. 

Finally, we investigate the multi-stage effect and the PBNE strategy pair for the long-term maximum utilities. To simplify the computation, we choose $c_2^0=2, c_2^1=0$, i.e., the expense of privilege escalations is more than lateral movements for the user and $c_1^0=0.1, c_1^1>2.39$.
Then, the optimal policy for both players is to choose NOP for all three expanded states at stage $t=1$. 
Therefore, although the attacker prefers to achieve a more advantageous system state $x^{t+1}$, aggressive behaviors $a_2^t=1$ at stage $t=1$ can decrease the defender's trust and result in a less favorable belief state $(\alpha^{t+1},\beta^{t+1})$. Because of the petty stage utility assumption, it is more beneficial for the attacker to remain stealthy at the intermediate stage and deceive the defender into a false belief. 
At the initial stage $t=0$, $P_1$ chooses $a_1^0=1$ with probability $0.20$. $P_2$ chooses $a_2^0=1$ when his type is larger than $0.95$ and chooses $a_2^0=0$ otherwise. The values are $V_1^0=59.97$ and $V_2^0=-10+13.57 \cdot \mathbf{1}_{\{\theta>0.95\}}$.
Therefore, it illustrates that the attacker of a large type value will take the risk of behaving aggressively to reach a desirable system state in the next stage because he would obtain higher  rewards once the attack succeeds. As a countermeasure, $P_1$ chooses to defend yet only with a small probability.

\section{Conclusion}\label{ConclusionSec}
In this work, we have explored a multistage incomplete information Bayesian game framework for designing proactive and adaptive defensive strategies for critical infrastructure networks with the presence of Advanced Persistent Threats (APTs). This framework well captures the multi-stage and multi-phase structure of APTs and their strategic nature to move stealthily within the network. 
With the information asymmetry between the attacker and the system, the defender needs to form a belief dynamically on the type of the user using observable footprints. To enable the online computation of the belief, we have used conjugate priors to reduce Bayesian updates into parameter updates. This approach leads to a computationally tractable extended-state dynamic programming criterion that yields an equilibrium solution consistent with the forward belief update and backward induction.
Finally, we have used Tennessee Eastman process as a case study to demonstrate the proposed framework. 
The numerical experiments have shown that our framework has significantly improved the security of critical infrastructures by strategically deterring the attacker and mitigating the APTs. 


\bibliographystyle{acm}
\bibliography{APT}


\end{document}